\documentstyle[psfig,epsf]{aipproc}
\def\lsim{\lower.5ex\hbox{$\; \buildrel < \over \sim \;$}}
\def\gsim{\lower.5ex\hbox{$\; \buildrel > \over \sim \;$}}
\input psfig.tex
\begin{document}
\title{On Radiative Acceleration of Jets and Outflows from Advective Disks}

\author{Indranil Chattopadhyay and Sandip K. 
Chakrabarti\thanks{Also, Honorary Scientist, Centre for 
Space Physics, IA-212, Salt Lake, Calcutta 700097}}
\address{S.N. Bose National Centre for Basic Sciences\\
JD Block, Salt Lake, Calcutta 700098\\}

\maketitle

\begin{abstract}
Jets and outflows are known to form out of advective accretion flows 
around black holes. Hard photons from the centrifugal barrier directly
hit the electrons and deposit momentum on them. For optically thick 
flows such deposition is not efficient, but for optically thin flows
matter could be accelerated to relativistic speed. In fact, even bound
matter could be made free through successive deposition. We 
discuss these possibilities.
\end{abstract}

\noindent Proceedings of Heidelberg International Gamma-Ray Astronomy Conference (Eds. F. Aharonian and H. Voelk)

\section*{Introduction}

The effect of radiative momentum deposition on astrophysical outflows 
around compact objects were studied using early disc models by many workers
(e.g. Vincent Icke~\cite{vi80} and Abramowicz and Piran~\cite{ap80}).
Chattopadhyay and Chakrabarti~\cite{1cc00,2cc00} ,
revisited the problem
using the radiation from advective discs~\cite{ct95}. 
The outflow was found to be accelerated appreciably, 
pushing bound matters
to infinity. 
Compact objects do not have atmospheres
like stars do,
therefore the outflows or jets 
have to be generated from the inflowing matter itself. 
As accreting  matter
comes closer to the compact object, it is slowed down by the centrifugal force 
and may cause shock~\cite{c89},
forcing the flow to jump discontinuously
from supersonic to subsonic state.
This hot, slowed
down region puffs up
in the form of a torus, known as the
centrifugal pressure supported boundary layer (hereafter, CENBOL).
Chakrabarti and Titarchuk~\cite{ct95} 
pointed out that radiation from this region may dictate whether a black
hole is in a soft state
or in a hard state. Similarly, Chakrabarti and co-workers~\cite{caa99,dc99} 
pointed out that the same region also causes formation of outflow.
In this region, the subsonic outflowing matter is
continuously bombarded by hot photons and hence apart from
usual thermal acceleration, radiative acceleration should be important
too. Thus the toroidal region or the CENBOL of accretion disc not only
emits high energy radiation, but its typical shape helps the
radiation to be focussed on the axis of symmetry. It is of no surprise
that we
find an impressive degree of acceleration at this region.
Our present exact
computation by radiative momentum deposition (RAMOD)
also vindicates earlier investigation~\cite{1cc00,2cc00} ,
that momentum deposition force 
can also drive bound matters as outflows. 
\section*{Calculation of RAMOD}
\noindent Solution of two-temperature equations inside a centrifugal
barrier supported shock
located at around $r=10 r_g$ ($1r_g=\frac{2GM_{B}}{c^2}$) is
presented in Chakrabarti and Titarchuk~\cite{ct95}. 
The inner surface of the CENBOL can be approximated by a cone, which we do
and compute RAMOD along the axis of symmetry.

\begin{figure}
\vbox{
\vskip -2.5cm
\hskip -0.0cm
\centerline{
\psfig{figure=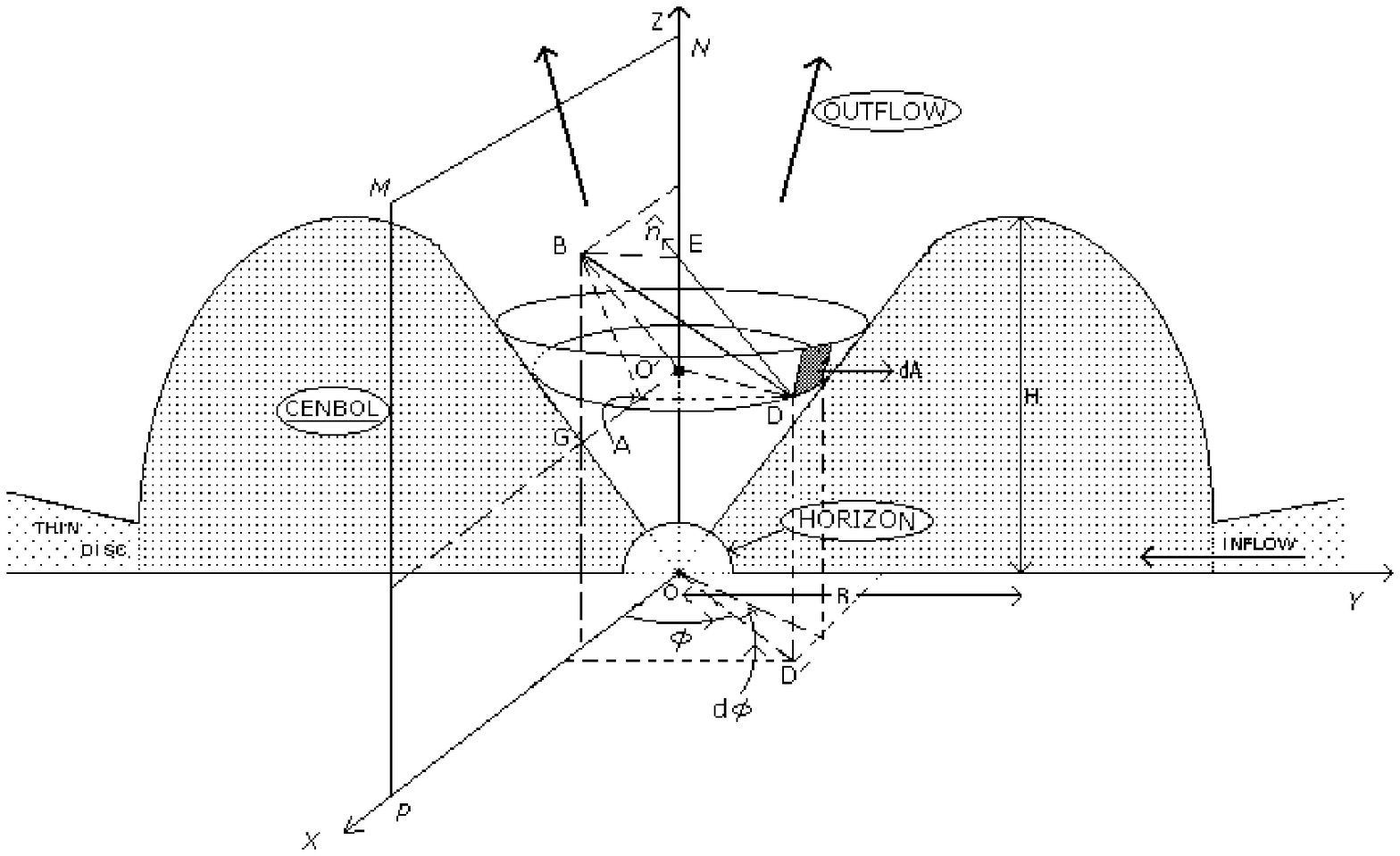,height=8truecm,width=9truecm}}}
\begin{verse}
\vspace{-2.5cm}
\noindent{\small{\bf Fig.1}:Schematic diagram of CENBOL
and disk/outflow geometry. $D$ is the source point, $B$ is the field point.
$O$ is the position
of compact object. The local unit normal $\hat{n}$ is along
$DC$. $OO^{'}=Z^{'}$, $O^{'}D=X$}
\end{verse}
\end{figure}

\noindent Figure 1, gives a schematic diagram of the disc/outflow geometry.
We choose cylindrical (r,$\phi$,z) coordinates, where the compact object is at 
the origin $O$. We assume the entire disc/outflow
system is symmetrical about the axis of rotation of the disc, therefore 
RAMOD has only 
two components ($D_r$ and $D_z$). 
We define a $\phi=0$ plane $PMNO$
The flux of radiation at field point $B(r,z)$i on $PMNO$, from differential
area $dA$ around source point $D(X, {\phi}, Z^{'})$ is given by:
$dF=I{\cos}{\angle (EDB)}d{\Omega}$,
where $I$ is the frequency averaged radiation intensity on the surface
and $d{\Omega}$ is the solid angle subtended by the differential area $dA$ at
$B$ and is equal to $dA/BD^2$.
The RAMOD will be along the flux i.e, along $BD$ and will 
have two components, one along the projection of $BD$ i.e, $AB$
and the other transverse to it. The transverse component will cancel each other
as we integrate on $\phi$.
\noindent Hence the two components of RAMOD is given by;
$$  
D_z(r,z)=D_0{\int}{\int}[{\frac{BD^2+ED^2-BE^2}{2ED}}]
{\frac{X}{\sin}{\theta}}
{\frac{BG}{BD^4}}d{\phi}dX
\eqno{(1)}
$$ 
\noindent and,
$$
D_r(r,z)=D_0{\int}{\int}[{\frac{BD^2+ED^2-BE^2}{2ED}}]
{\frac{X}{\sin}{\theta}}
{\frac{AG}{BD^4}}d{\phi}dX
\eqno{(2)}
$$
\noindent where, $D_0=\frac{2I{\sigma}_T}{cm_p}$, $\theta=$semi vertical
angle of the inner surface of CENBOL, $m_p=$mass of the proton, $c=$ 
velocity of light. 
To estimate 
$I$ we assume that the dissipation of
binding energy of the last stable orbit of accreting
matter is the source of radiation. Hence,
$I=\frac{L}{{\Omega}_{T}{\cal A}}$,
where $L={\eta}{{\dot M}_{acc}}c^2$, is the amount of rest mass
energy of the accreting matter converted into radiation per unit
time. Here ${\dot M}_{acc}$ is the accretion rate. ${\eta}$
is the convertion ratio and is equal to $0.06$ for the
Schwarzschild metric.
${\Omega}_{T}=2{\pi}$, is the solid angle in which the radiation is
coming out locally.
${\cal A}=2{\pi}R(H^2+R^2)^{1/2}$, is the total area of conical
CENBOL (see, Fig. 1).
In our calculation of RAMOD we do not consider
the attenuation of the radiation field.
Optically thick flow
would be difficult to accelerate this way. 
\noindent Our assumption of optically thin flow
(i.e., $\tau <1$) puts limits on the mass outflow rate
and the mass of the central object.
\noindent The density of the radial outflow is given by;
$$
\rho=\frac{\dot M}{v{\it r}^2},
\eqno{(3a)}
$$
\noindent where ${\it r}=$radius vector of outflow in spherical
geometry, $v=$radial velocity. 
Let ${\it r}=L=10r_g$  be the length scale where
the above approximation is valid. 
We rewrite the expression of the density using the definition of sound speed,
$$
\rho=\frac{{\dot M}{\mu}^{1/2}{m_p}^{1/2}}{{L}^2{\gamma}^{1/2}{k_B}^{1/2}
T^{1/2}}.
\eqno{(3b)}
$$
In a radiation dominated flow, $P_{gas}<<P_{rad}$ i.e,
$$
\rho<<\frac{{\sigma}T^3{\mu}{m_p}}{k_B},
\eqno{(4)}
$$
which provides a upper limit of density. Combining
expressions (3b) and (4), we get;
$$
\rho{\approx}{\left[\frac{{\dot M}^6{\sigma}{\mu}^4{m_p}^4}
{{\it L}^{12}{\gamma}^3{k_B}^4}\right]^{1/7}}.
\eqno{(5)}
$$

\noindent The condition for the optically thin flow ($\tau<1$)
in the Thompson scattering regime
for length scale ${\sim} 10 r_g$, also provides the estimate of upper limit of 
density,
$$
\rho < {\frac{1.69{\times}10^{27}}{M_B}} {\rm gm\ cm^{-3}}.
\eqno{(6)}
$$
From (5) and (6), 
in order
to have optically thin outflows, we get:
$$
\frac{{\dot M}^6}{{M_B}^5}<1.7{\times}10^{-70} {\rm gm\ sec^{-6}}
\eqno{(7)}
$$
or,
$$
{{\dot m}_{10}}^6m_{10}<6.1{\times}10^{-8},
\eqno{(8)}
$$
where, ${\dot m}_{10}$ is the mass outflow rate ($\dot M$)
in the units of Eddington rate for a $10 M_{\odot}$ black hole
and $m_{10}$ is the black hole
mass ($M_B$) in units of $10 M_{\odot}$. 
Increasing ${\dot M}_{acc}$ increases disk luminosity
and hence  magnitude of RAMOD. However, Chakrabarti~\cite{c97} showed
that higher ${\dot M}_{acc}$ results in higher outflow rate ${\dot M}$,
which results in outflows with higher density $\rho$ and therefore
higher $\tau$. In case of optically thick outflows
radiative acceleration is not efficient,
as the intensity of radiation from the disc will be attenuated
by a factor $e^{-{\tau}}$ while passing through the flow.
In our work the outflow considered are optically thin and hence
respects equations (7) and (8).
On integrating equations (1) and (2) from 
${\phi}=0{\rightarrow}2{\pi}$
and $X=1.5r_g{\rightarrow}R$, we get the value of RAMOD everywhere.
Figure 2, shows the vector field of RAMOD. We choose ${\dot M}_{acc}=
10{\dot M}_{EDD}$. We have also chosen the shock location
for accretion to be at a distance $10 r_g$~\cite{ct95} on the equatorial
plane of the accretion disc,
the specific angular momentum of the accreting matter to be
$1.6 \frac{2GM_{B}}{c}$, and thus we have $H=4.4 r_g$, $R=6.671 r_g$.

\begin{figure}
\vbox{
\vskip -2.0cm
\hskip 4.0cm
\psfig{figure=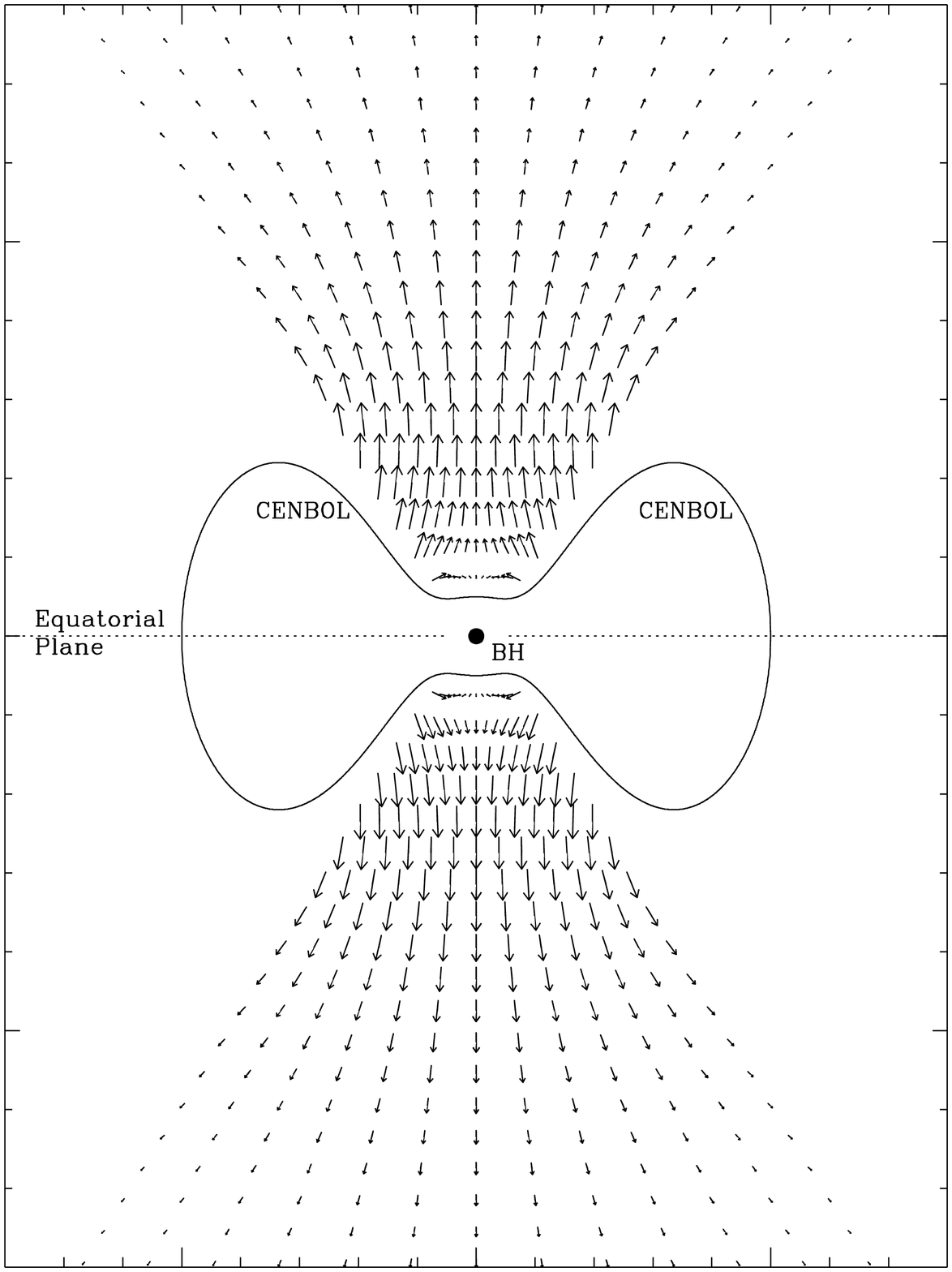,height=6truecm,width=6truecm,angle=0}}
\begin{verse}
\vspace{0.0cm}
\noindent{\small{\bf Fig.2}: The vector field of radiative force 
due to the radiation from CENBOL. ${\dot M}_{acc}=10{\dot M}_{EDD}$.
The position of shock $10 r_g$, specific  angular momentum of the
accreting matter $1.6 \frac{2GM_{B}}{c}$, $H=4.4 r_g$, $R=6.671 r_g$.}
\end{verse}
\end{figure}

\noindent From Fig. 2 we see that very close to the horizon the
radiation force is directed towards it. As one goes away, 
RAMOD becomes positive it is still directed towards
the axis of symmetry. Further out it falls off
${\sim} 1/r^2$ and force field becomes more and more radial.
This 
 focusing effect,
will tend to collimate the outflow.

\section*{Global solution with radiative momentum deposition.}
\noindent As a special case of solutions of equations (1) and (2), 
RAMOD along the axis of symmetry i.e. $D_z(0,z)$,
was calculated by Chattopadhyay
and Chakrabarti~\cite{2cc00} and was approximated
as $D_{\it r} {\sim} D_z(0,z)$, for conical outflow very close to
the axis of symmetry.

\begin{figure}
\vbox{
\vskip -5.0cm
\hskip 2.5cm
\psfig{figure=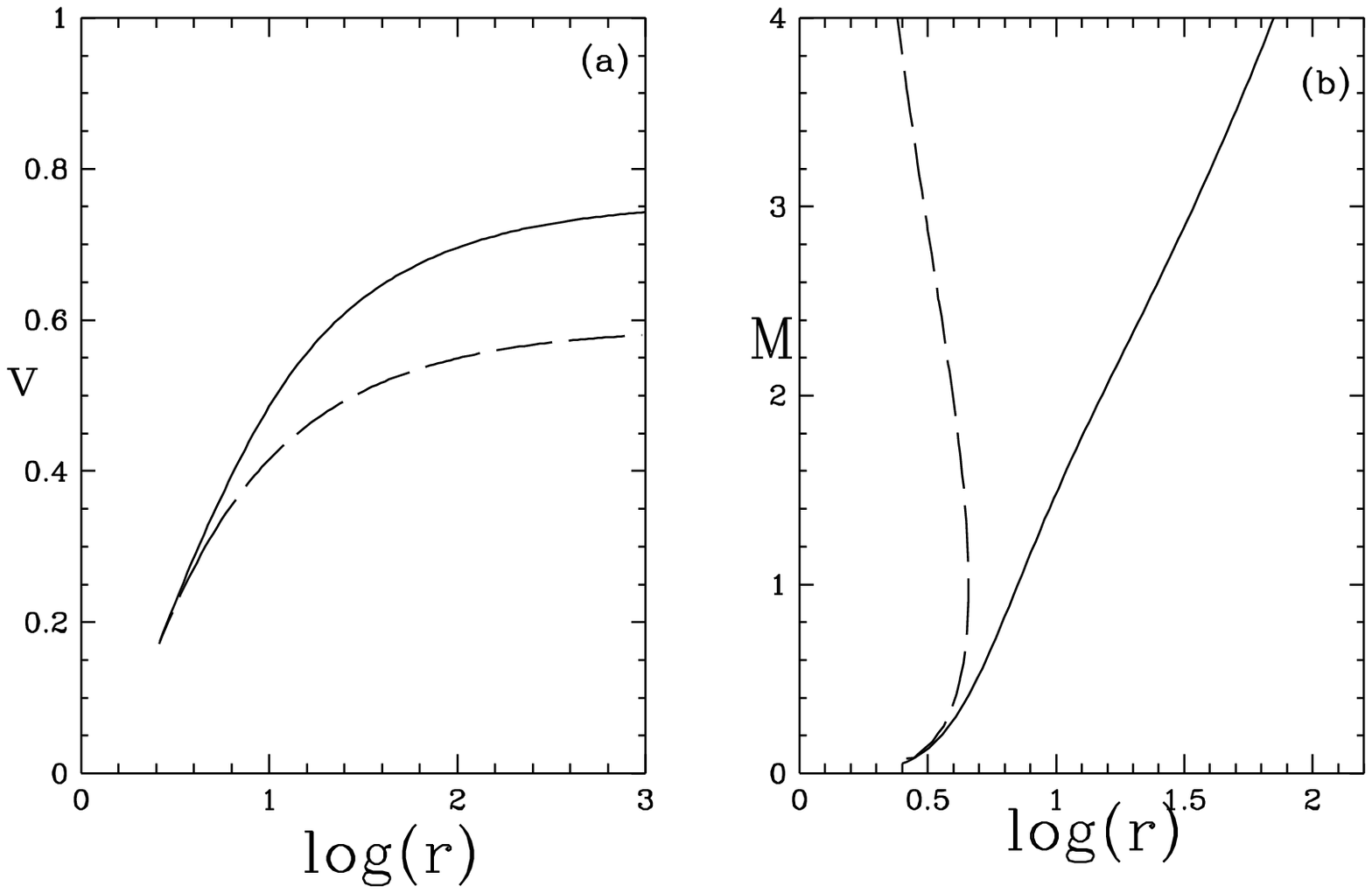,height=9truecm,width=10truecm,angle=0}}
\begin{verse}
\vspace{-2.0cm}
\noindent{\small{\bf Fig.3(a-b)}:Variation with $log(r)$ of (a)
$v$, RAMOD corresponding to $10{\dot M}_{EDD}$ (solid) and Bondi
outflow (long dashed); (b)Mach number $M$, RAMOD corresponds
to  ${\dot M}_{acc}=15{\dot M}_{EDD}$ (solid) and thermally
driven outflow (long-dashed). RAMOD solution becomes unbound.}
\end{verse}
\end{figure}

$$
{\cal D}_{\it r}={\it D}_0
[{\cos}2({\Phi}_{1}-{\theta})-{\cos}2({\Phi}_{2}
-{\theta})]
\eqno{(9)}
$$
\noindent where,
${\Phi}_{1}=tan^{-1}\left [\frac{2(H-r{\cos}^{2}
{\theta})}{r{\sin}2{\theta}}\right ] $ 
and,
${\Phi}_{2}=tan^{-1}\left [\frac{2(1-r{\cos}^{2}
{\theta})}{r{\sin}2{\theta}}\right ] $ \\
\noindent $H$ is the maximum height of the CENBOL from the
equatorial plane and, ${\it D}_0={\frac{{\pi}I{\sigma_{T}}}
{2cm_{p}}} $,
where $I$, ${\sigma_{T}}$, $c$, $m_{p}$ has there usual meaning.
Using expression (9) in the momentum balance equation, and in the entropy 
generation equation we solve for flow topology of the radial outflow.
Figure 3a shows the comparison of radial velocity $v$ with and without
the RAMOD. Velocity is plotted against  $log(r)$.  The solid line
corresponds to $10 {\dot M}_{EDD}$ and
the long dashed line is the solution of Bondi~\cite{hb52} outflow, 
i.e. outflow
not acted on by RAMOD. The initial velocity at which flow is launched
for the two cases is  $0.1717 c$ at radial distance $2.598 r_g$.
The $v_t$ for the above cases are $0.743 c$,
$0.5804 c$ respectively. 
We see that RAMOD accelerates the outflow
and higher RAMOD results higher terminal speed.
RAMOD not only accelerates but also increases the energy of the flow
often freeing bound
matter. Figure 3b shows the comparison of 
Mach number $M (=v/a)$ variation variation with ${\rm log(r)}$.
The long-dashed curve represents
the thermally driven negative energy flow, and
the solid curve represent the outflow which is acted on by
RAMOD corresponding to ${\dot M}_{acc} = 15 {\dot M}_{EDD}$.
The initial parameter for both is, $E_{in}=-0.0349 c^2$
at $r_{in}=2.509 r_g$. In this case of later,
RAMOD
increases the energy of the flow to positive
value, pushing it to infinity as free outflow and making it
supersonic in the  process, while the former remains
bound and ultimately dives back to the compact object.

This work is partly supported by DST project Analytical and Numerical
Studies of Astrophysical Flows Around Compact Objects.


\begin{references}
\bibitem{vi80} Icke, V. {\it Astron. J.}, {\bf 85}(3) 329  (1980).
\bibitem{ap80} Abramowicz, M. A. and Piran, T. {it Astrophys. Journ.} {\bf 241} L7 (1980).
\bibitem{1cc00}  Chattopadhyay, I. and Chakrabarti, S. K.  {\it Int. Journ. Mod. Phys. D}, 
 {\bf 9}(1) 57 (2000).
\bibitem{2cc00} Chattopadhyay, I. and Chakrabarti, S. K.  {\it Int. Journ. Mod. Phys. D},
 {\it in press}.
\bibitem{ct95} Chakrabarti S.K. and Titarchuk, L.G.  {\it Astrophys. \ J.}\ {\bf 455}, 623 (1995).
\bibitem{c89}  Chakrabarti, S. K. {\it Astrophys. J.}, {\bf 347} 365 (1989).
\bibitem{caa99} Chakrabarti, S. K. {\it Astron. Astrophys.}, {\bf 351} 185 (1999).
\bibitem{dc99}  Das T.K. and  Chakrabarti S.K. {\it Class.\ Quant.\ Grav.}\ {\bf 16}, 3879 (1999).
\bibitem{c97} Chakrabarti, S. K. {\it Astrophys. J.}, {\bf 484} 313 (1997)
\bibitem{hb52} Bondi, H. {\it Mon. Not. Roy. Soc}, {\bf 112} 195 (1952)

\end{references}
\end{document}